\documentclass[twocolumn,english]{revtex4}
\usepackage[LGR,LGR,T1]{fontenc}
\usepackage[latin2]{inputenc}
\usepackage[letterpaper]{geometry}
\usepackage{textcomp}
\usepackage{amsmath}
\usepackage{amssymb}
\usepackage{graphicx}

\makeatletter

\DeclareRobustCommand{\greektext}{%
  \fontencoding{LGR}\selectfont\def\encodingdefault{LGR}}
\DeclareRobustCommand{\textgreek}[1]{\leavevmode{\greektext #1}}
\ProvideTextCommand{\~}{LGR}[1]{\char126#1}

\@ifundefined{textcolor}{}
{%
 \definecolor{BLACK}{gray}{0}
 \definecolor{WHITE}{gray}{1}
 \definecolor{RED}{rgb}{1,0,0}
 \definecolor{GREEN}{rgb}{0,1,0}
 \definecolor{BLUE}{rgb}{0,0,1}
 \definecolor{CYAN}{cmyk}{1,0,0,0}
 \definecolor{MAGENTA}{cmyk}{0,1,0,0}
 \definecolor{YELLOW}{cmyk}{0,0,1,0}
}

\usepackage{babel}

\usepackage{babel}

\makeatother

\usepackage{babel}
\begin{document}

\title{{\LARGE{}{}{}Robust field-dressed spectra of diatomics in an optical
lattice}}

\author{Mariusz Pawlak$^{1}$, Tamás Szidarovszky$^{2}$, Gábor J. Halász$^{3}$
and Ágnes Vibók$^{4,5}$ }

\address{{\large{}{}{}$^{1}$}Faculty of Chemistry, Nicolaus Copernicus
University in Toruń, Gagarina 7, 87-100 Toruń, Poland }

\address{{\large{}{}{}$^{2}$}Laboratory of Molecular Structure and Dynamics,
Institute of Chemistry, ELTE Eötvös Loránd University and MTA-ELTE
Complex Chemical Systems Research Group, Pázmány Péter sétány 1/A,
H-1117 Budapest, Hungary }

\address{{\large{}{}{}$^{3}$}Department of Information Technology, University
of Debrecen, H-4002 Debrecen, PO Box 400, Hungary}

\address{{\large{}{}{}$^{4}$}Department of Theoretical Physics, University
of Debrecen, H-4002 Debrecen, PO Box 400, Hungary }

\address{{\large{}{}{}$^{5}$}ELI-ALPS, ELI-HU Non-Profit Ltd, H-6720 Szeged,
Dugonics tér 13, Hungary}
\begin{abstract}
The absorption spectra of the cold $\mathrm{Na_{2}}$ molecule dressed
by a linearly polarized standing laser wave is investigated. In the
studied scenario the rotational motion of the molecules is frozen
while the vibrational and translational degrees of freedom are accounted
for as dynamical variables. In such a situation a light-induced conical
intersection (LICI) can be formed. To measure the spectra a weak field
is used whose propagation direction is perpendicular to the direction
of the dressing field but has identical polarization direction. Although
LICIs are present in our model, the simulations demonstrate a very
robust absorption spectrum, which is insensitive to the intensity
and the wavelength of the dressing field and which does not reflect
clear signatures of light-induced nonadiabatic phenomena related to
the strong mixing between the electronic, vibration and translational
motions. However, by widening artificially the very narrow translational
energy level gaps, the fingerprint of the LICI appears to some extent
in the spectrum.
\end{abstract}
\maketitle

\section{Introduction}

Nonadiabatic processes associated with avoided crossings (ACs) or
conical intersections (CIs) \cite{Neumann,Lenz1,Yarkony,Baer1,Graham1,Domcke,Baer2}
always involve nuclear dynamics on at least two coupled potential
energy surfaces (PESs) and thus can not be treated within the Born-Oppenheimer
(BO) approximation \cite{Born-Oppenheimer}. In this case the electronic
states are coupled by the nuclear motion, and the energy exchange
between the fast electrons and the slow moving nuclei may become significant.
CIs between electronic PESs play a key mechanistic role in chemistry,
physics and biology, most notably in photochemistry and photobiology.
In several important cases like fragmentation, charge transfer, isomerization
processes of polyatomics or radiationless relaxation of the excited
electronic states \cite{martin,kukura,Hahn,Schultz,Lan,Olivucci,Tamas2},
the CIs can provide very efficient decay channel on femtosecond time
scale. CIs show up in different contexts as well, such as in optical
\cite{Born,Berry,Peleg} and solid state physics \cite{Simon,Levy}.
Furthermore, exceptional points \textendash{} degeneracies between
continuum states \textendash{} can also be found in nature \cite{Lenz2,Lenz3}. 

CIs can exist only if the molecule possesses at least two independent
nuclear vibrational coordinates. In field free diatomics that have
only one nuclear degree of freedom CIs are never formed. However,
CIs can occur in triatomic systems and they are fairly abundant in
truly large polyatomic molecules.

If the laser field is present, then due to the interaction of the
diatomics with the field, the rotational nuclear coordinate comes
into play and serves as an additional degree of freedom. CIs can be
created both by running \cite{Lenz4} and by standing laser waves
\cite{Lenz5} even in diatomic molecules. In the former situation
the laser field rotates the molecule due to the interaction of the
transition dipole moment of the molecule with the electric field which
leads to an effective torque toward the polarization direction of
the light and so called ``light-induced conical intersections''
(LICIs) are formed. The position of these LICIs are determined by
the laser frequency while the laser intensity controls the strength
of the nonadiabatic coupling. Numerous theoretical \cite{Gabi1,Andris1}
and experimental \cite{exp1,exp2} studies have confirmed that the
LICIs have strong impact on the dynamical \cite{Gabi1,Andris1} and
spectroscopic \cite{Lenz4,Tamas1} properties of diatomics. Recently
the frequency and intensity dependence of the field-dressed rovibronic
spectrum of the diatomic Na$_{2}$ molecule in gas phase was extensively
discussed \cite{Tamas1,Tamas_PRA}.

Light-induced nonadiabatic phenomena can also occur in optical lattices,
which have been widely applied in atomic and diatomic physics so as
to cool, trap and control different properties of atoms and molecules
\cite{Dominic}. The optical lattice is a periodic potential energy
landscape that the molecules experience as a result of the standing
wave pattern created by the interference of two counter propagating
laser beams. The nonadiabatic phenomenon can emerge when the translational
motion of the molecule strongly couples with the electronic, vibrational
and rotational motions leading to a periodic array of light-induced
CIs. In this case the translational motion provides the new dynamical
variable which is necessary to form a branching space for diatomics.
Historically, the LICIs have been predicted first in optical lattices
for the $\mathrm{Na_{2}}$ molecule \cite{Lenz5}. It was demonstrated
that the trapping efficiency of the $\mathrm{Na_{2}}$ molecule in
its lowest electronic state can be significantly reduced due to the
presence of the LICI \cite{Lenz5}. Later on the intensively studied
cold weakly bound rubidium molecule \cite{Bendkowsky,Koch} has been
considered which plays an important role in the cold-atom physics.
As a results, due to the effect of the LICI a significant localization
enhancement of the $\mathrm{Rb_{2}}$ has been revealed \cite{Pawlak}.
Besides the theoretical studies important experimental papers have
also discussed the nonadiabatic effects \cite{Tanya0} and the spectra
of ultracold molecules in an optical lattice \cite{Tanya1,Tanya2}.

In the present work we focus again on the two-atom homonuclear cold
sodium molecule and discuss the standing field-dressed spectra of
this object. Our aims are two-fold. First, by investigating the frequency
and intensity dependence of the spectra, we would like to explore
to what extent the spectra of a non-rotating $\mathrm{Na_{2}}$ dressed
by an optical lattice and a freely rotating $\mathrm{Na_{2}}$ dressed
by a running laser wave are similar or different. Second, how the
impact of the light-induced nonadiabatic phenomena can be visualized
on the absorption spectra of the standing field-dressed non-rotating
$\mathrm{Na_{2}}$ molecule. 

The article is arranged as follows. In the next section, we present
a brief outline of the theory and the computation details. In the
third section results are shown and analyzed. Finally, conclusions
are given in the last section.

\section{Theory and computation details}

In the present paper a two step numerical pump-probe simulation is
derived to study the weak-field absorption spectrum of the field-dressed
homonuclear $\mathrm{Na_{2}}$ molecule in an optical lattice.

\textit{\emph{First, we specify the field-dressed states of the system
by applying a medium intensity ``pump'' }}standing laser wave\textit{\emph{
which mixes the field-free eigenstates of the sodium dimer. }}The
applied standing laser wave can be written as the sum of two counter
propagating linearly polarized running waves \cite{nimrod}: 
\begin{align}
\varepsilon_{0}\left[\sin\left(k_{L}Z+\omega_{L}t\right)+\sin\left(k_{L}Z-\omega_{L}t\right)\right]/2 & =\label{eq:standing-wave-1}\\
=\varepsilon_{0}\sin\left(k_{L}Z\right)\cos\left(\omega_{L}t\right)\nonumber 
\end{align}
where $\varepsilon_{0}$ is the amplitude, $\omega_{L}$ is the frequency
and $k_{L}$ is the wave vector ($k_{L}=$$\omega_{L}/c=2\pi/\lambda_{L}$)
of the laser field.\textit{\emph{ }}

\textit{\emph{Second, we determine the dipole transition amplitudes
between the field-dressed states in the framework of first-order time-dependent
perturbation theory by using a second weak probe pulse.}}\textit{
} 
\begin{figure}
\begin{centering}
\includegraphics[width=0.5\textwidth]{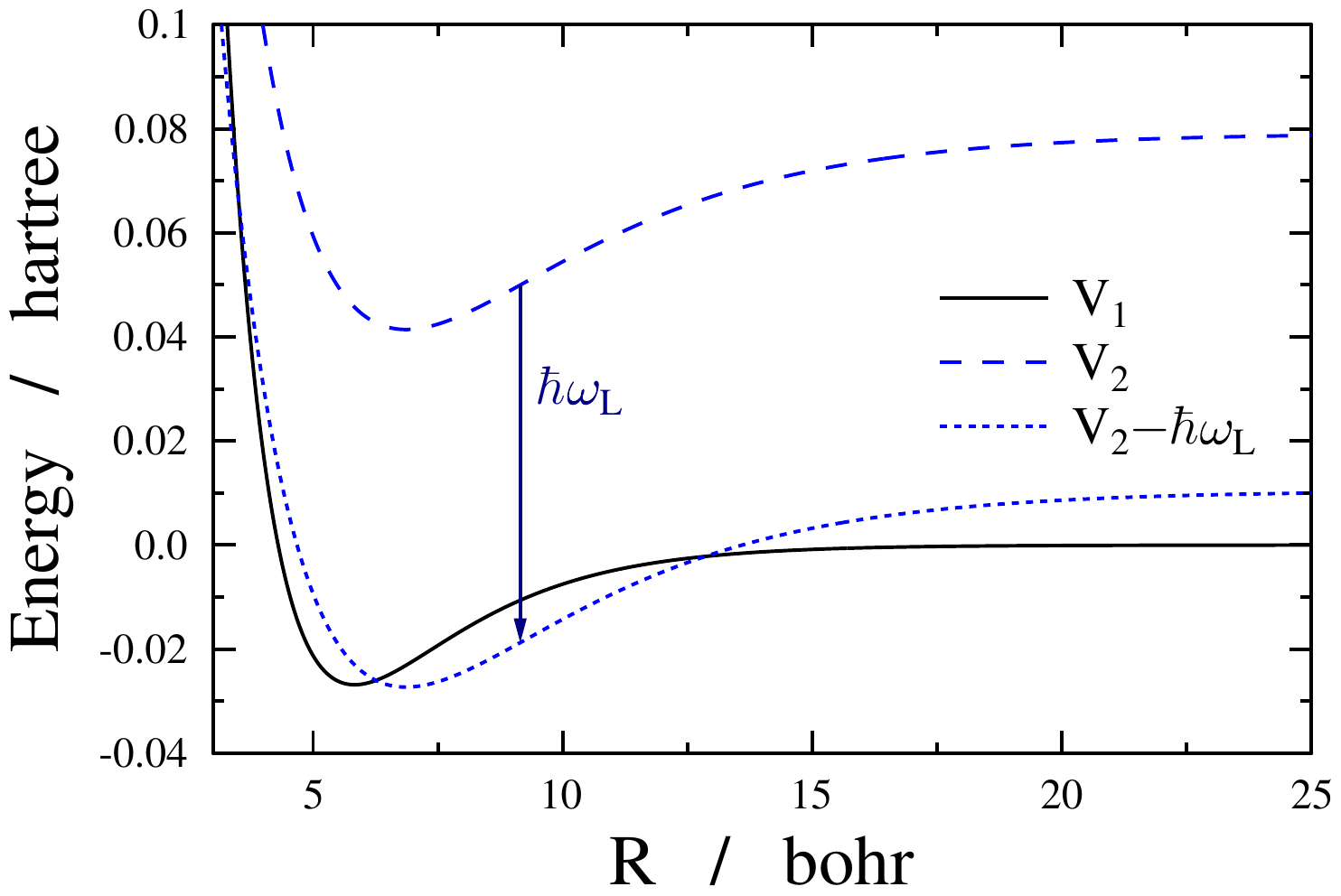}\\
 \includegraphics[width=0.5\textwidth]{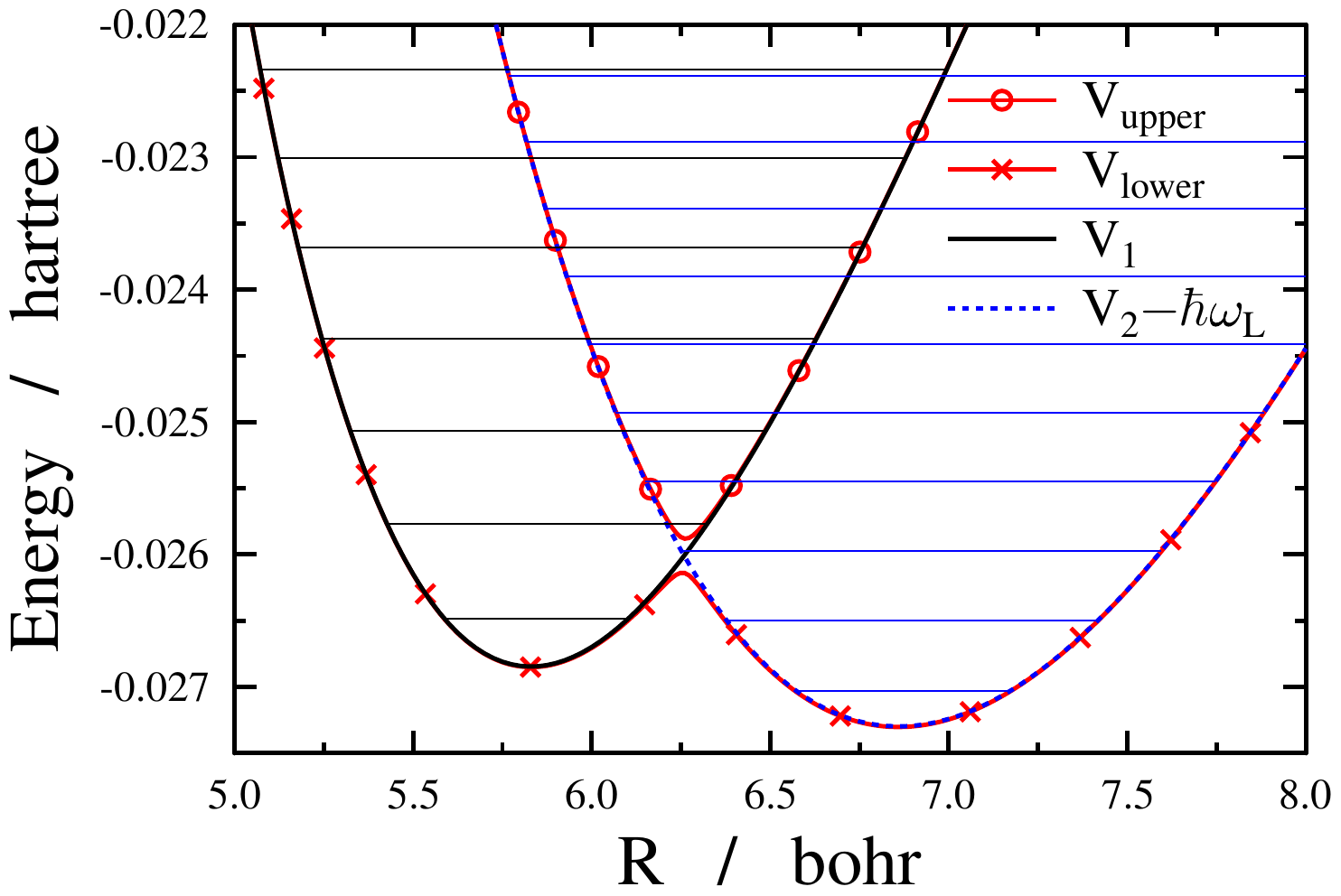}\\
 \includegraphics[width=0.5\textwidth]{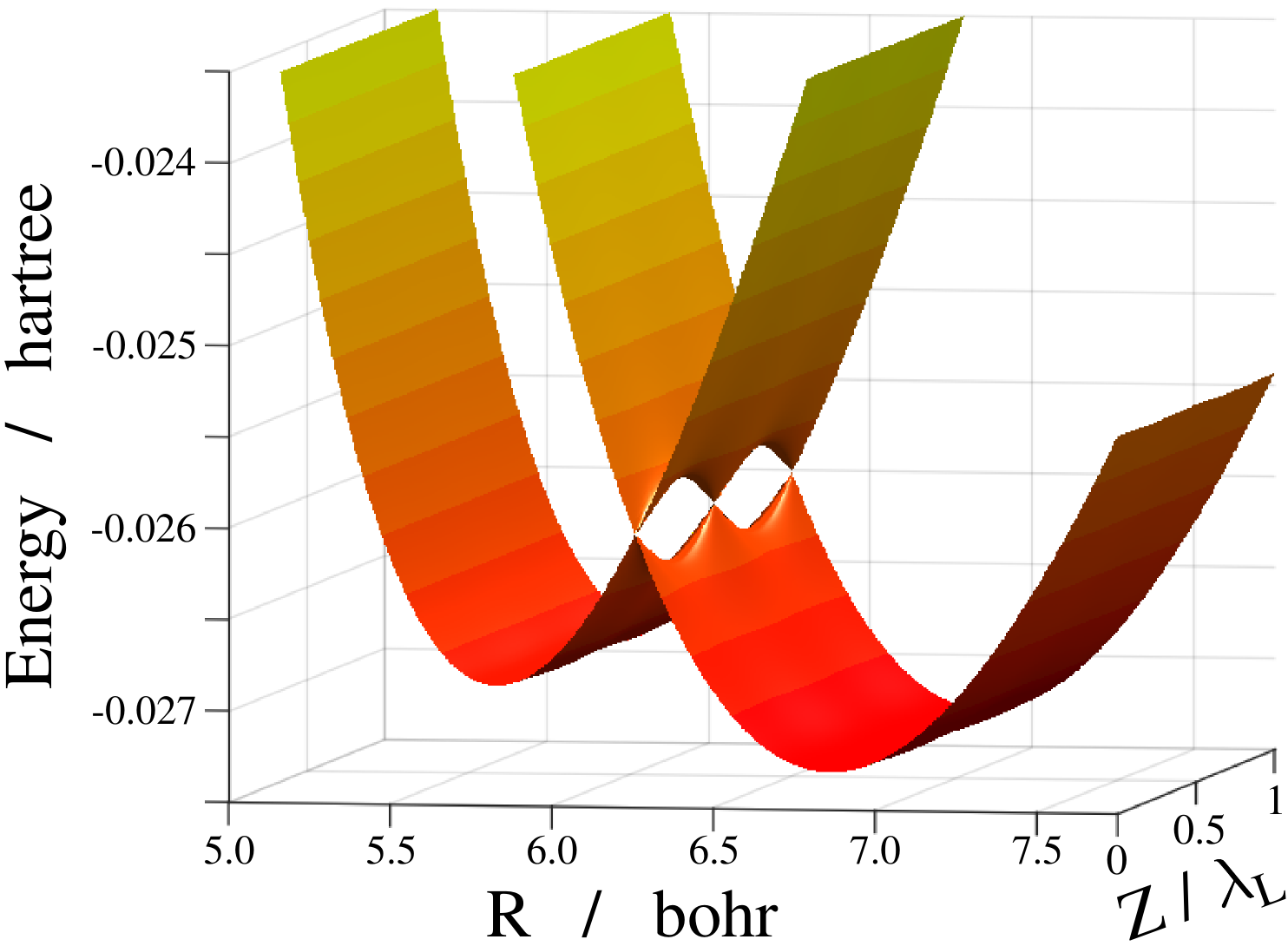} 
\par\end{centering}
\caption{\label{Pot} Potential energies of the $\mathrm{Na_{2}}$ molecule.
Upper panel: potential energy curves V$_{1}$ and V$_{2}$ associated
with the $X^{1}\Sigma_{g}^{+}$ and $A^{1}\Sigma_{u}^{+}$ electronic
states of Na$_{2}$. V$_{2}$ is dressed by laser light applying $\hbar\omega_{L}$
dressed photon energy. Middle panel: diabatic (V$_{1}$, V$_{2}$
- $\hbar\omega_{L}$) and adiabatic (V$_{\textrm{upper}},$ V$_{l\textrm{ower}}$)
one dimensional cuts of the potential energy surfaces along the interatomic
separation. Laser intensity and wavelength are 5 \texttimes{} 10$^{8}$
W/cm$^{2}$ and \textgreek{l}$_{L}$ = 663 nm, respectively. The different
horizontal lines provide the appropriate diabatic vibrational energy
levels. Lower panel: periodic array of light-induced conical intersection
(LICI) points between the V$_{\textrm{lower}}$ and V$_{\textrm{upper}}$
adiabatic potential energy surfaces. The applied field intensity and
wavelength are I$_{0}=$5 \texttimes{} 10$^{8}$ W/cm$^{2}$ and \textgreek{l}$_{L}$
= 663 nm, respectively. }
\end{figure}

The two relevant electronic states of the $\mathrm{Na_{2}}$ molecule
which are involved in the numerical calculations are the ground $X^{1}\Sigma_{g}^{+}$
and the first excited $A^{1}\Sigma_{u}^{+}$ states (further denoted
as $\psi_{1}^{e}$ and $\psi_{2}^{e}$, respectively). These are characterized
by the $V_{1}(R)$ and $V_{2}(R)$ potential energy curves (see Fig.~\ref{Pot}).

We represent the time-dependent Hamiltonian in the Floquet form \cite{Floquet}.

Assuming dipole approximation and only one net photon absorption the
well known ($2\times2$) static dressed state form can be obtained
for the Hamiltonian \footnote{In the dressed state representation the interaction between the molecule
and the electromagnetic field is obtained by shifting the energy of
the appropriate electronic states by $n\hbar\omega_{L}$ ($n=1,2,...$$\infty)$.
As one net photon absorption is assumed, $n=1.$}:

\begin{align}
\mathbf{H} & =\left(\begin{array}{cc}
\hat{T}_{R\theta\phi}+\hat{T}_{XYZ} & 0\\
0 & \hat{T}_{R\theta\phi}+\hat{T}_{XYZ}
\end{array}\right)\label{eq:Hamiltonian}\\
 & +\left(\begin{array}{cc}
V_{1}(R) & \frac{1}{2}\varepsilon_{0}d(R)\cos\theta\sin(k_{L}Z)\\
\frac{1}{2}\varepsilon_{0}d(R)\cos\theta\sin(k_{L}Z) & V_{2}(R)-\hbar\omega_{L}
\end{array}\right).\nonumber 
\end{align}
Here $\hat{T}_{R\theta\phi}$ and $\hat{T}_{XYZ}$ are the kinetic
energy operators of the molecular rovibrational motion and of the
translational motion of the center of mass of the molecule, respectively.
These are written in the following form: 
\begin{equation}
\hat{T}_{R,\theta,\phi}=-\frac{\hbar^{2}}{2\mu}\frac{\partial^{2}}{\partial R^{2}}+\frac{\hat{L}^{2}(\theta,\phi)}{2\mu R^{2}},\label{eq:rovibronic}
\end{equation}
\begin{equation}
\hat{T}_{XYZ}=-\frac{\hbar^{2}}{2M}\left(\frac{\partial^{2}}{\partial X^{2}}+\frac{\partial^{2}}{\partial Y^{2}}+\frac{\partial^{2}}{\partial Z^{2}}\right),\label{eq:translational-energy}
\end{equation}
where $\mu$ is the reduced and $M$ is the total mass of the two
atoms, respectively, $\hat{L}^{2}(\theta,\phi)$ is the squared angular
momentum operator associated with diatomic rotations, $R$ is the
molecular vibrational coordinate, and $d(R)=-\left\langle \psi_{1}^{e}\left|\sum_{j}r_{j}\right|\psi_{2}^{e}\right\rangle $
is the transition dipole moment which is parallel to the molecular
axis. $\theta$ denotes the angle between the polarization direction
and the direction of the transition dipole and thus one of the angles
of rotation of the molecule. The light is propagated along the $Z$
direction. The potentials $V_{1}(R)$ and $V_{2}(R)$ were taken from
Ref.~\cite{Magnier}, whereas the transition dipole moment from Ref.~\cite{Zemke}.

The form of the Hamiltonian demonstrated in Eq.~(\ref{eq:Hamiltonian})
helps to understand the essence of the light-induced conical intersection
phenomenon. In this representation the laser light shifts the $V_{2}(R)$
excited potential curve downwards by $\hbar\omega_{L}$ (see the upper
panel of Fig.~\ref{Pot}) and a crossing is formed between the $V_{1}(R)$
diabatic ground and the ($V_{2}(R)-\hbar\omega_{L}$) shifted diabatic
excited potential energy curves (see the middle panel of Fig.~\ref{Pot}).
After diagonalizing the potential energy matrix the adiabatic potential
curves $V_{{\rm {lower}}}$ and $V_{{\rm {upper}}}$ can be obtained
(middle panel of Fig.~\ref{Pot}). These two curves can cross each
other, giving rise to a light-induced conical intersection whenever
the following two conditions are simultaneously fulfilled: (\emph{i})
$V_{1}(R)=V_{2}(R)-\hbar\omega_{L}$; ($ii$) $Z=n\lambda_{L}/2$
($n$ is an integer) or $\theta=\pi/2$. The lower panel of Fig.~\ref{Pot}
visualizes CIs induced by a laser light for the sodium dimer.

Now, we reduce the three-dimensional problem to a two-dimensional
one by assuming that the initial ground electronic and rotational
state $\psi_{1}^{e}(J=0)$ is coupled with only the first excited
electronic and rotational state $\psi_{2}^{e}(J=1)$, $\langle\psi_{1}^{e}(J=0)|\cos\theta|\psi_{2}^{e}(J=1)\rangle=\frac{1}{\sqrt{3}}$.
Then, we can obtain the field-dressed (FD) eigenstates $\Psi_{q}^{{\rm {FD}}}(R,Z)$
and the corresponding quasi-energies $E_{q}^{{\rm {FD}}}(R,Z)$ by
solving the matrix eigenvalue problem with the Hamiltonian of Eq.~(\ref{eq:Hamiltonian}).
Therefore, let us build up the Hamiltonian matrix by using basis functions
which are the products of plain wave functions describing the translational
motion of the system and field-free molecular vibrational eigenfunctions
of the ground and first excited electronic states:
\begin{equation}
\varphi_{\nu}^{(j)}(R)\Phi_{k}(Z).\label{eq:basis}
\end{equation}
Here 
\begin{equation}
\varphi_{\nu}^{(j)}(R)=\sum_{\eta}C_{\nu,\eta}^{(j)}\psi_{\eta}(R)\label{eq:basis1}
\end{equation}
is the $\nu$th vibrational eigenfunction of the $j$th electronic
state, where $j=1$ and $j=2$ denote the $X^{1}\Sigma_{g}^{+}$ and
$A^{1}\Sigma_{u}^{+}$ electronic states, respectively. The $\psi_{\eta}(R)$
functions in Eq. (\ref{eq:basis1}) stand for the basis function used
for expanding the vibrational states. Moreover 
\begin{equation}
\Phi_{k}(Z)=\sqrt{\frac{1}{L}}\exp\left(\frac{ik2\pi}{L}Z\right),\quad k=-N,...,N\label{eq:basis3}
\end{equation}
describes the plain wave functions moving along $Z$ direction, where
$i$ is the imaginary unit, $L=\lambda_{L}$, and $N=100$.

The resulting field-dressed states in the Floquet framework can be
obtained as the linear combination of the products of our basis functions,
given spherical harmonics, and the Fourier vectors of the Floquet
states, namely 
\begin{eqnarray}
|\Psi_{q}^{{\rm FD}}\rangle & = & \sum_{\nu,k}c_{q;\nu,k}^{(1)}|\varphi_{\nu}^{(1)}(R)\rangle|\Phi_{k}(Z)\rangle|Y_{0,0}(\theta,\phi)\rangle|n\rangle\nonumber \\
 & + & \sum_{\nu,k}c_{q;\nu,k}^{(2)}|\varphi_{\nu}^{(2)}(R)\rangle|\Phi_{k}(Z)\rangle|Y_{1,0}(\theta,\phi)\rangle|n-1\rangle,\nonumber \\
\label{eq:field-dressed}
\end{eqnarray}
where $c_{q;\nu,k}^{(j)}$ are the expansion coefficients gained by
diagonalizing the Hamiltonian of Eq.~(\ref{eq:Hamiltonian}) after
representing it in the basis set of Eq.~(\ref{eq:basis}).

Using the field-dressed states of Eq.~(\ref{eq:field-dressed}) one
can determine the spectrum of the field-dressed molecule. We assume
a weak probe pulse whose propagation direction is perpendicular to
the direction of the pump pulse but has identical polarization direction.
Furthermore, we assume that the transitions induced by this probe
pulse can be described by one-photon processes. In this case the transition
amplitudes in dipole approximation can be computed in the framework
of first-order time-dependent perturbation theory (TDPT) which is
a standard procedure of computational molecular spectroscopy \cite{Spect1}.
Then, the transition amplitude from a $\vert\Psi_{q^{\prime}}^{{\rm FD}}\rangle$
initial state to a $\vert\Psi_{q}^{{\rm FD}}\rangle$ final state
reads

\begin{widetext} 
\begin{eqnarray}
 &  & \left\langle \Psi_{q^{\prime}}^{{\rm FD}}\left|\hat{\mathbf{d}}\hat{\mathbf{e}}\right|\Psi_{q}^{{\rm FD}}\right\rangle =\left\langle \Psi_{q^{\prime}}^{{\rm FD}}(R,\theta,\phi,Z)\left|d(R)\cos\theta\right|\Psi_{q}^{{\rm FD}}(R,\theta,\phi,Z)\right\rangle =\nonumber \\
 & = & \sum_{\nu^{\prime},k^{\prime}}\sum_{\nu,k}\left(c_{q^{\prime};\nu^{\prime},k^{\prime}}^{(1)}\right)^{\ast}c_{q;\nu,k}^{(2)}\langle\varphi_{\nu^{\prime}}^{(1)}(R)|d(R)|\varphi_{\nu}^{(2)}(R)\rangle\langle\Phi_{k^{\prime}}(Z)|\Phi_{k}(Z)\rangle\langle Y_{0,0}(\theta,\phi)|\cos\theta|Y_{1,0}(\theta,\phi)\rangle\langle n^{\prime}|n-1\rangle\nonumber \\
 &  & +\sum_{\nu^{\prime},k^{\prime}}\sum_{\nu,k}\left(c_{q^{\prime};\nu^{\prime},k^{\prime}}^{(2)}\right)^{\ast}c_{q;\nu,k}^{(1)}\langle\varphi_{\nu^{\prime}}^{(2)}(R)|d(R)|\varphi_{\nu}^{(1)}(R)\rangle\langle\Phi_{k^{\prime}}(Z)|\Phi_{k}(Z)\rangle\langle Y_{1,0}(\theta,\phi)|\cos\theta|Y_{0,0}(\theta,\phi)\rangle\langle n^{\prime}-1|n\rangle\nonumber \\
 & = & \frac{1}{\sqrt{3}}\delta_{n^{\prime},n-1}\sum_{\nu^{\prime},k^{\prime}}\sum_{\nu,k}\left(c_{q^{\prime};\nu^{\prime},k^{\prime}}^{(1)}\right)^{\ast}c_{q;\nu,k}^{(2)}\delta_{k^{\prime},k}\sum_{\eta^{\prime}}\sum_{\eta}\left(C_{\nu^{\prime},\eta^{\prime}}^{(1)}\right)^{\ast}C_{\nu,\eta}^{(2)}\langle\psi_{\eta^{\prime}}^{(1)}(R)|d(R)|\psi_{\eta}^{(2)}(R)\rangle\nonumber \\
 &  & +\frac{1}{\sqrt{3}}\delta_{n^{\prime}-1,n}\sum_{\nu^{\prime},k^{\prime}}\sum_{\nu,k}\left(c_{q^{\prime};\nu^{\prime},k^{\prime}}^{(2)}\right)^{\ast}c_{q;\nu,k}^{(1)}\delta_{k^{\prime},k}\sum_{\eta^{\prime}}\sum_{\eta}\left(C_{\nu^{\prime},\eta^{\prime}}^{(2)}\right)^{\ast}C_{\nu,\eta}^{(1)}\langle\psi_{\eta^{\prime}}^{(2)}(R)|d(R)|\psi_{\eta}^{(1)}(R)\rangle\label{eq:transition}
\end{eqnarray}
\end{widetext} Here, the $\hat{\mathbf{e}}$ unit vector defines
the polarization direction of the dressing pump and probe pulses.
The first term of the expression stands for absorption while the second
term represents the simulated emission spectra. From now on we will
focus only on the absorption, therefore the first term will serve
as our working formula.

\begin{figure}
\includegraphics[width=0.48\textwidth]{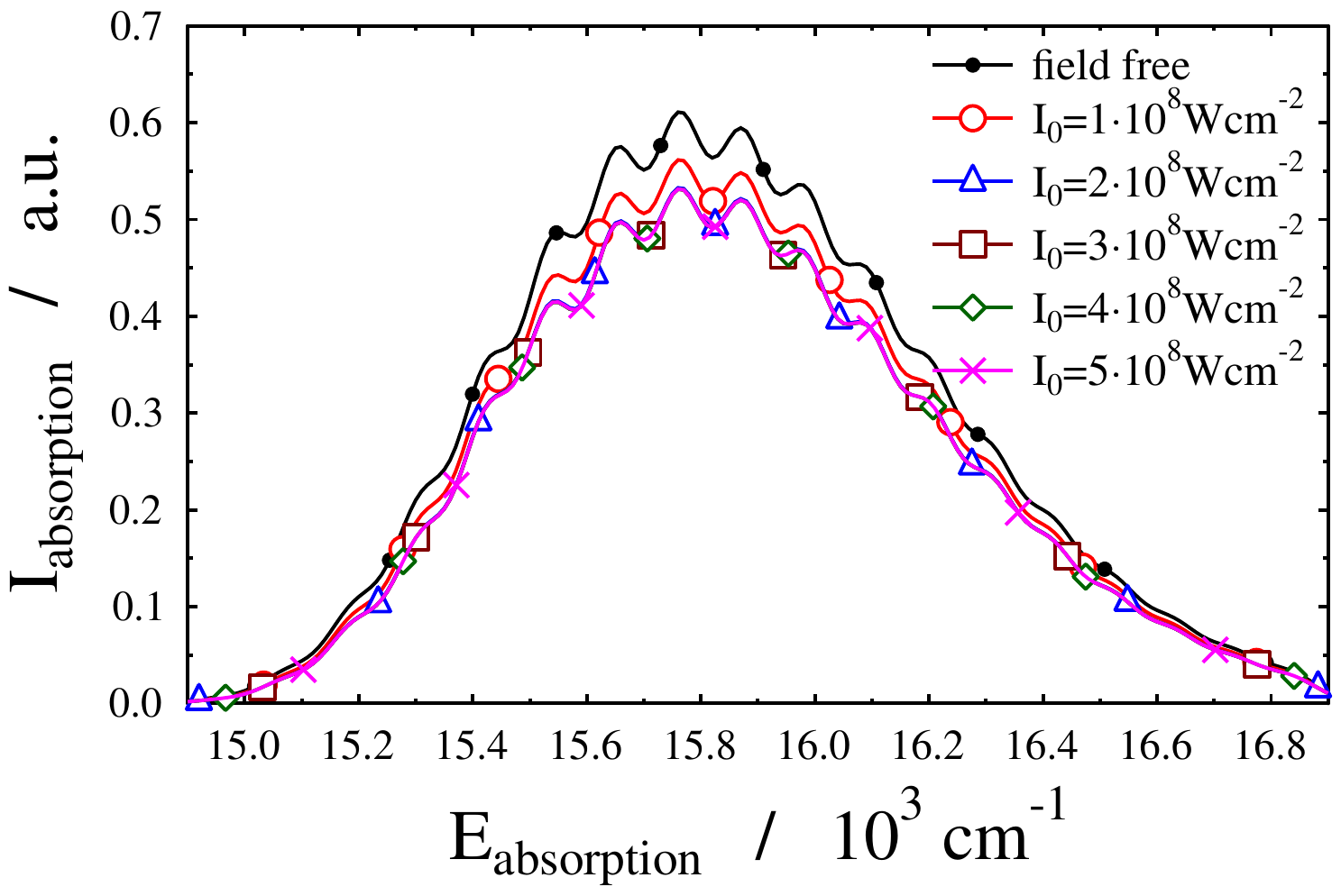}

\includegraphics[width=0.48\textwidth]{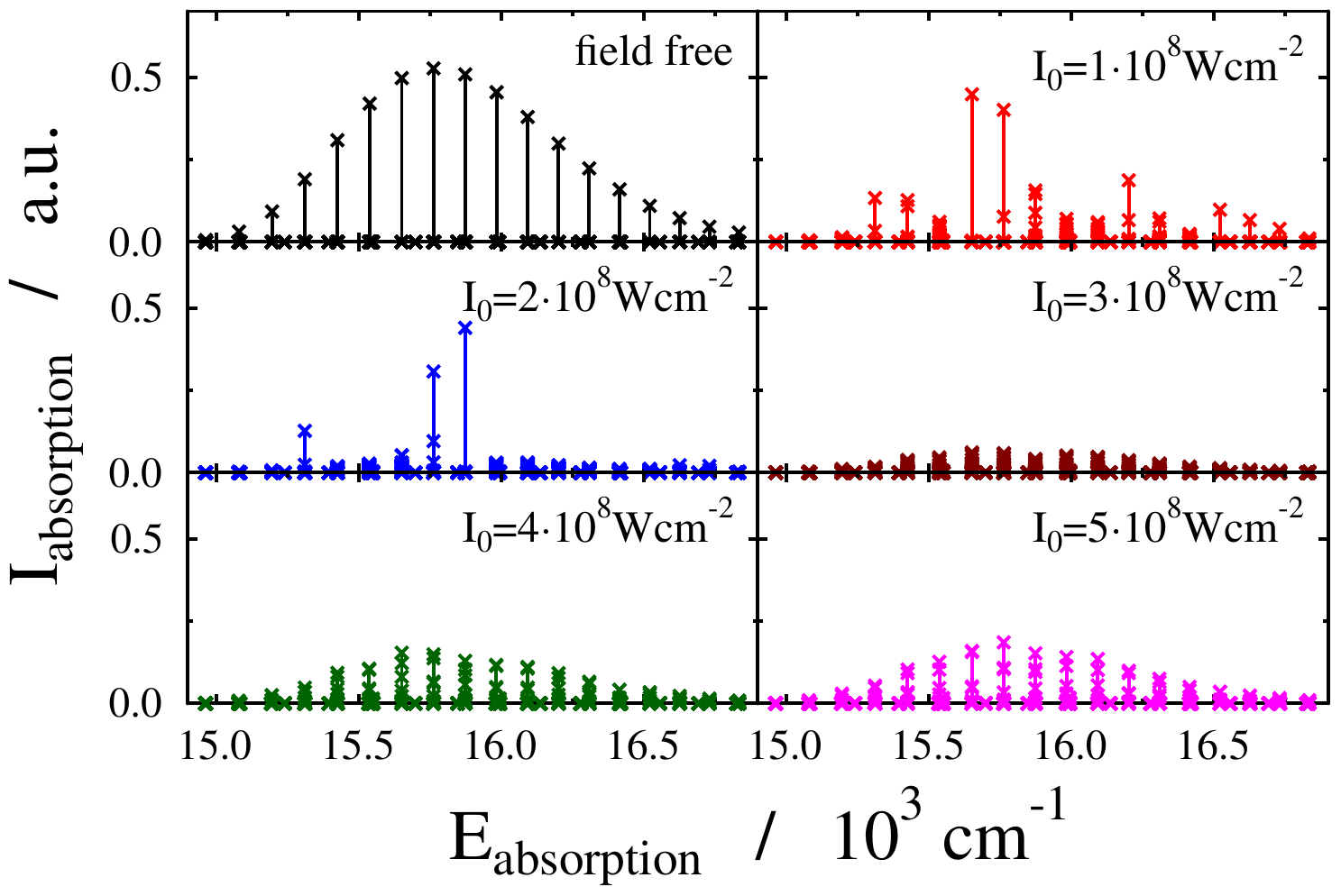}

\caption{\label{fig:2} Absorption spectra of the standing field-dressed Na$_{2}$
as a function of dressing-field (\textgreek{l}$_{L}$ = 663 nm) intensity,
computed using the first expression of Eq (\ref{eq:transition}).
The spectra only show transitions from a single field-dressed state,
which was chosen by ``adiabatic approximation''. The curves shown
in the upper panel are obtained by convolving the stick spectra in
the lower panel with a Gaussian function having \textgreek{sv} = 50
cm$^{-1}$. }
\end{figure}

\begin{figure}
\includegraphics[width=0.48\textwidth]{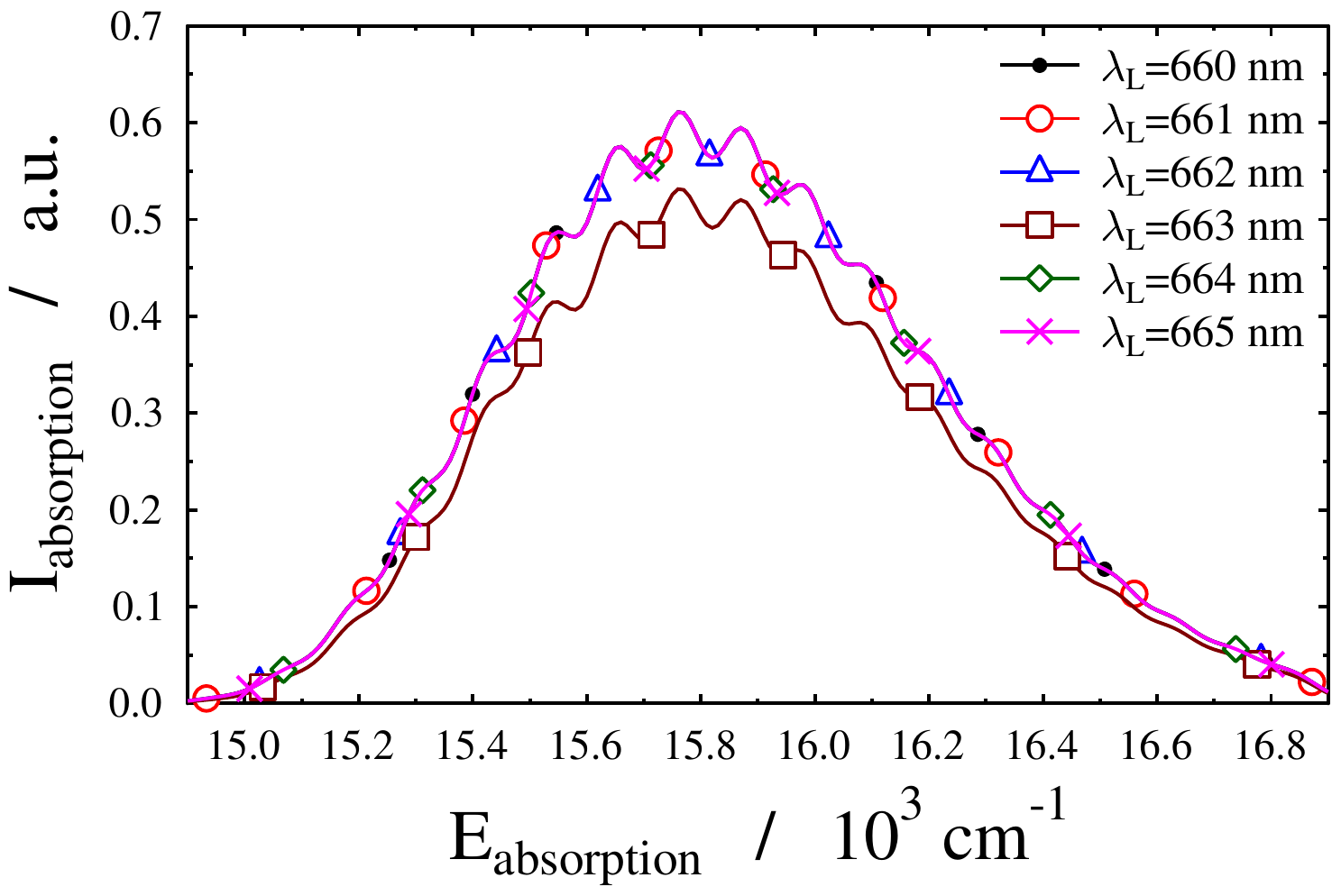}

\caption{\label{fig:3} Absorption spectra of the standing field-dressed Na$_{2}$
at six different dressing-field wavelengths and a dressing-field intensity
of I$_{0}$ =5 \texttimes{} 10$^{8}$ W/cm$^{2}$. The spectra only
show transitions from a single field-dressed state, which was chosen
by ``adiabatic approximation''.}
\end{figure}

\begin{figure}
\includegraphics[width=0.48\textwidth]{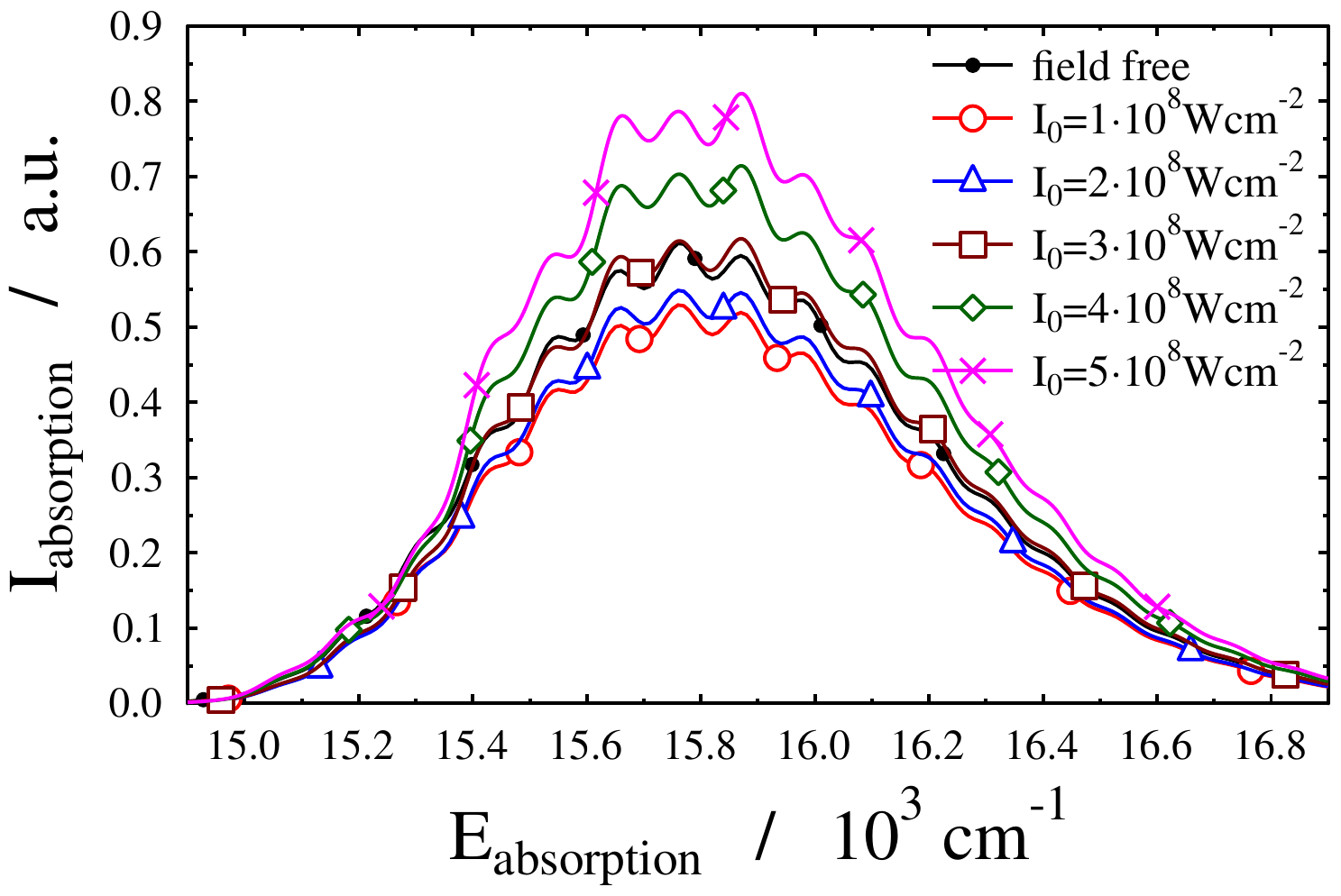}

\includegraphics[width=0.48\textwidth]{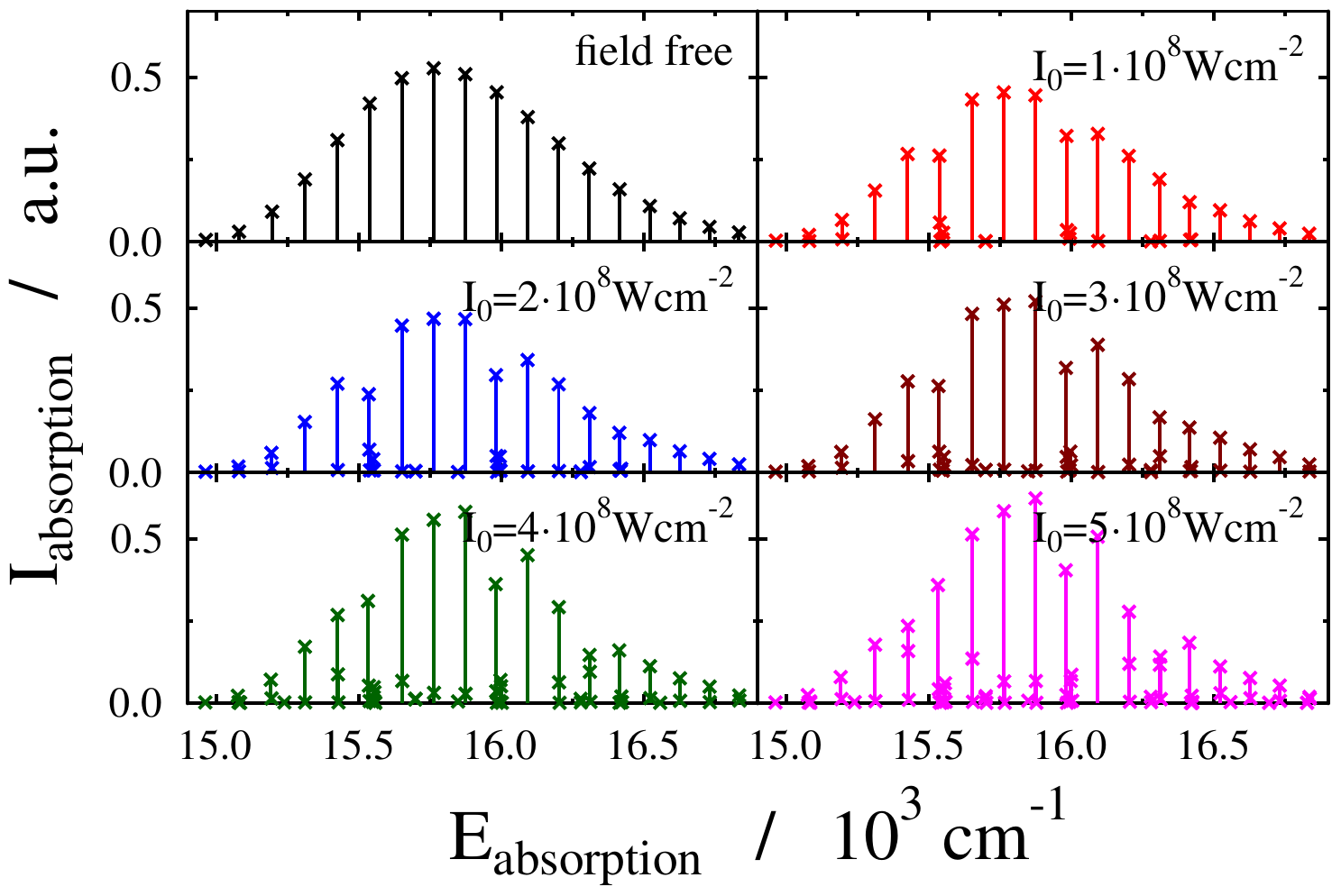}

\caption{\label{fig:4}Absorption spectra of the running field-dressed Na$_{2}$
as a function of dressing-field (\textgreek{l}$_{L}$ = 663 nm) intensity,
computed using the first expression of the last line in Eq. (3) of
\cite{Tamas1}. The spectra only show transitions from a single field-dressed
state, which was chosen by ``adiabatic approximation''. The curves
shown in the upper panel are obtained by convolving the stick spectra
in the lower panel with a Gaussian function having \textgreek{sv}
= 50 cm$^{-1}$. }
\end{figure}

\begin{figure}
\includegraphics[width=0.48\textwidth]{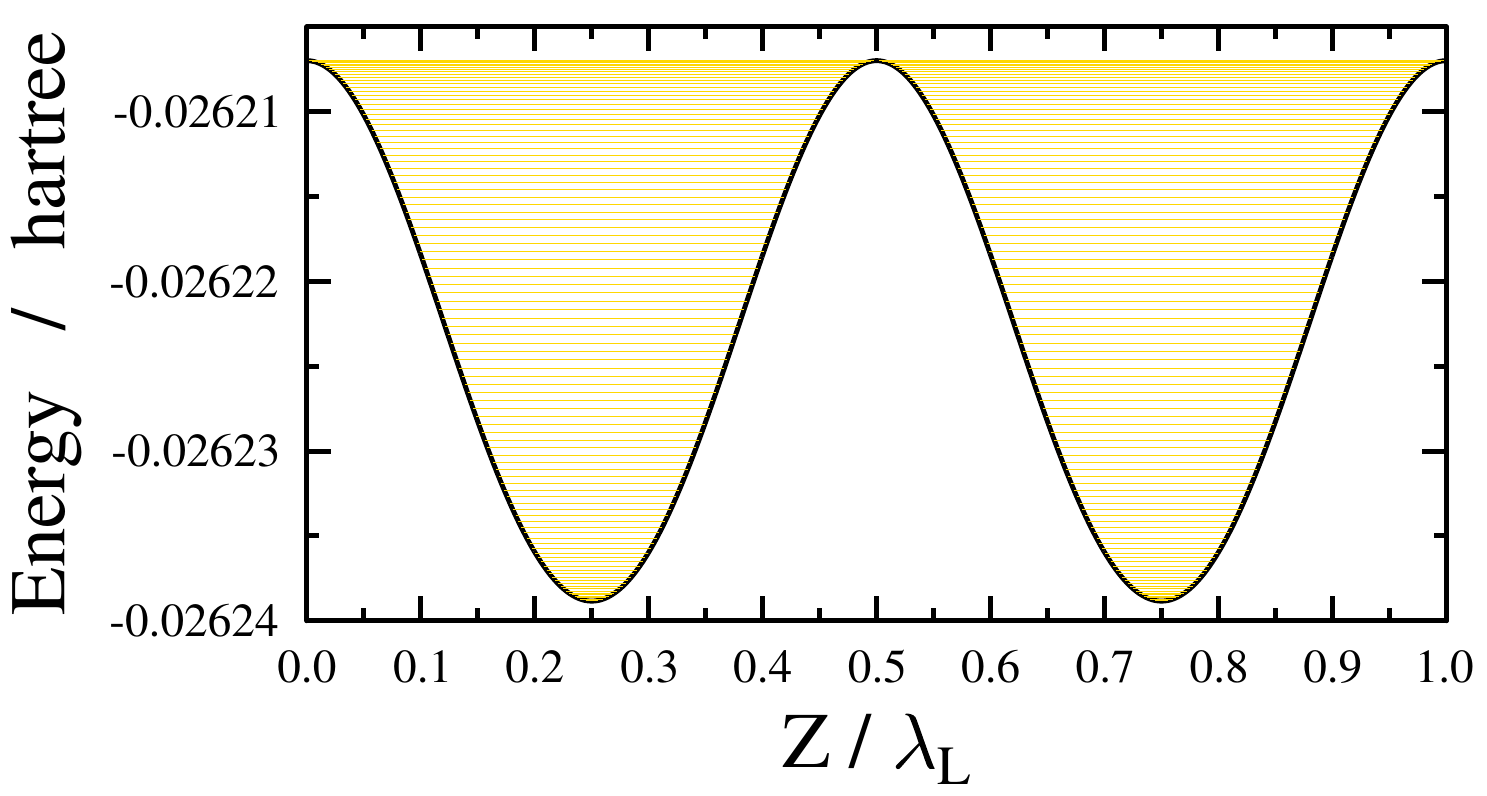}

\includegraphics[width=0.48\textwidth]{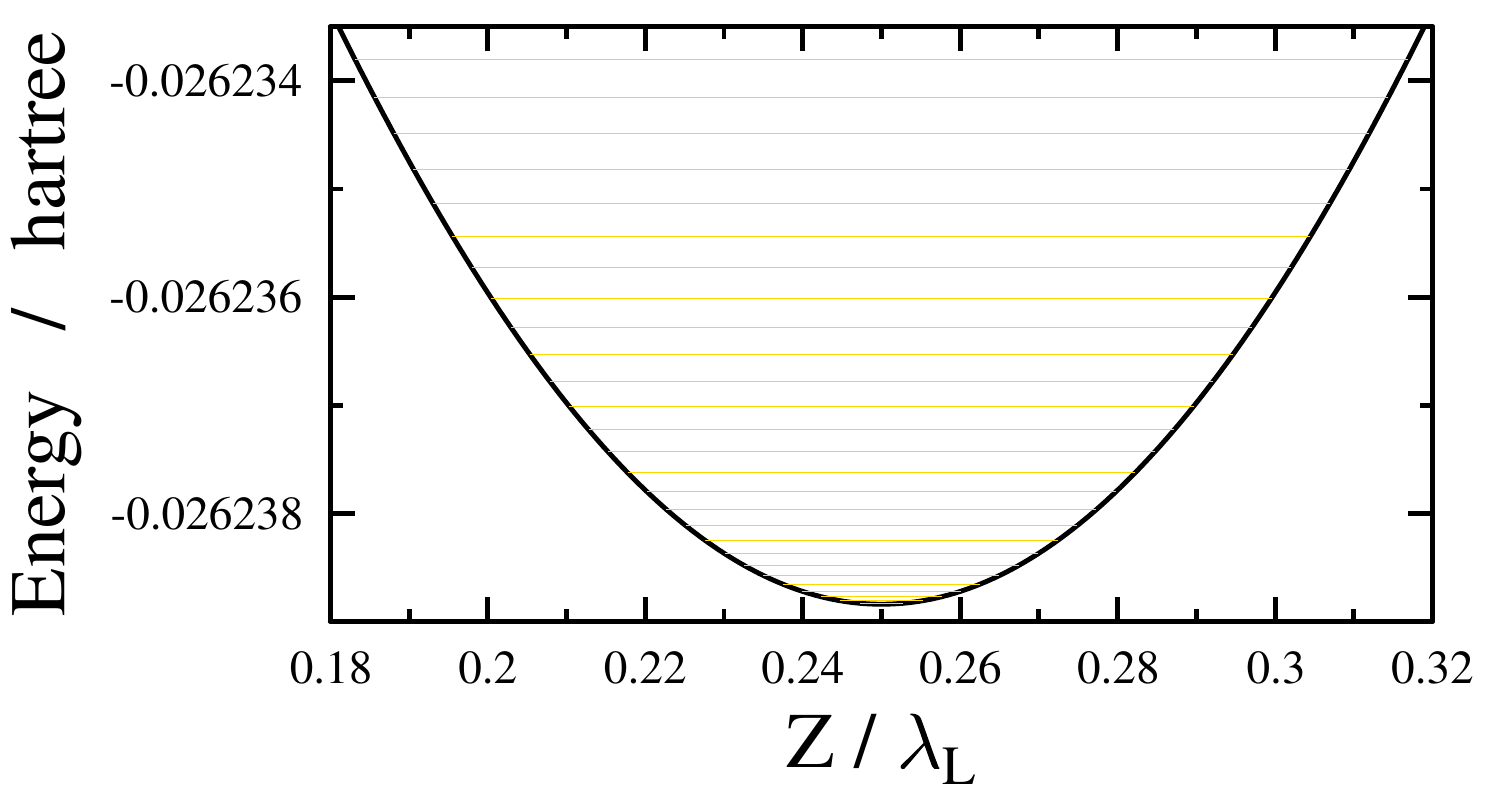}

\caption{\label{fig:5}One dimensional cuts ($R=6.2$ a.u.) of the translational
energy levels of the Na$_{2}$ molecule in an optical lattice. The
dressing-field wavelength and intensity are \textgreek{l}$_{L}$ =
663 nm and I$_{0}$ =5 \texttimes{} 10$^{8}$ W/cm$^{2}$, respectively.}
\end{figure}

\begin{figure*}
\includegraphics[width=0.48\textwidth]{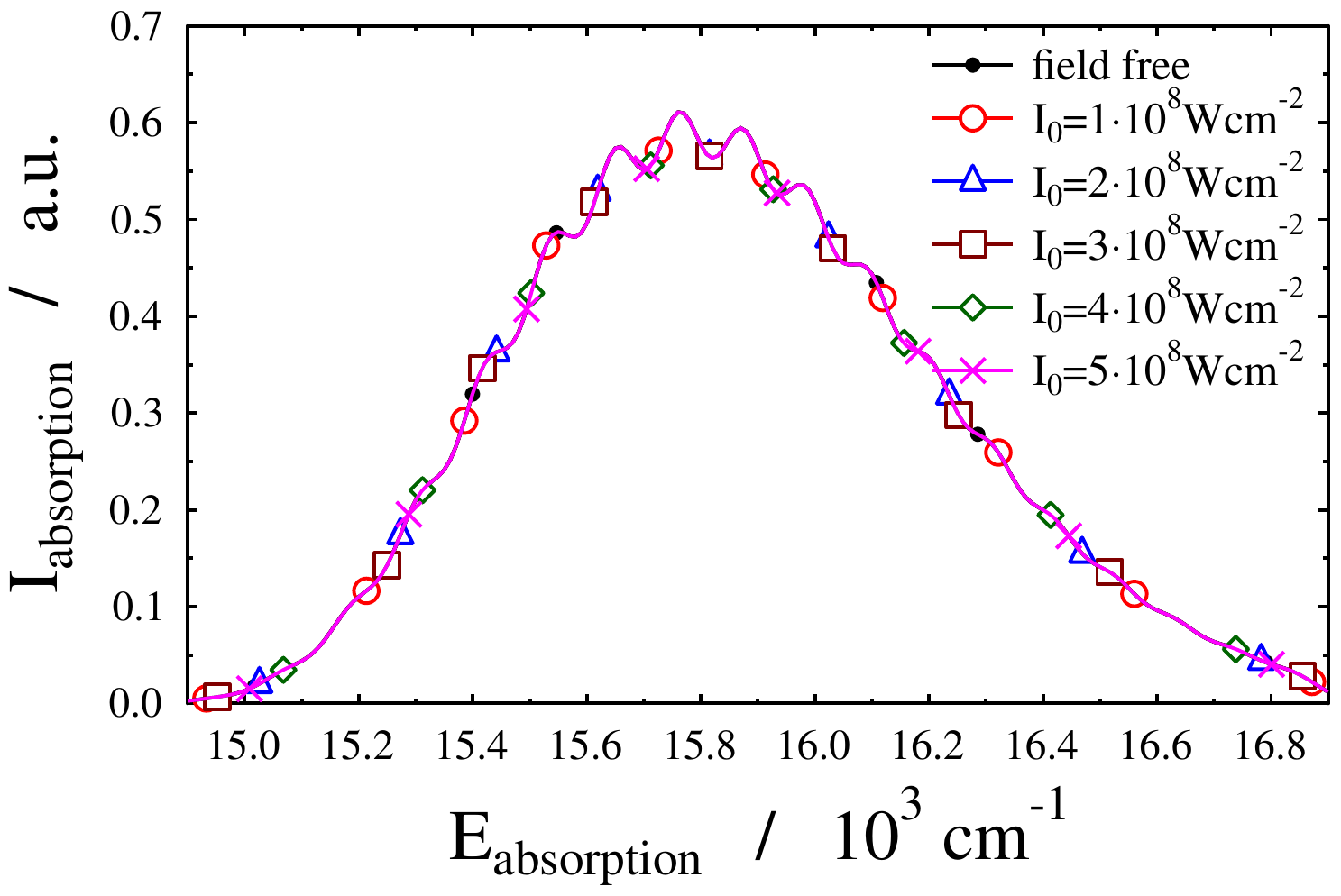}\hfill{}\includegraphics[width=0.48\textwidth]{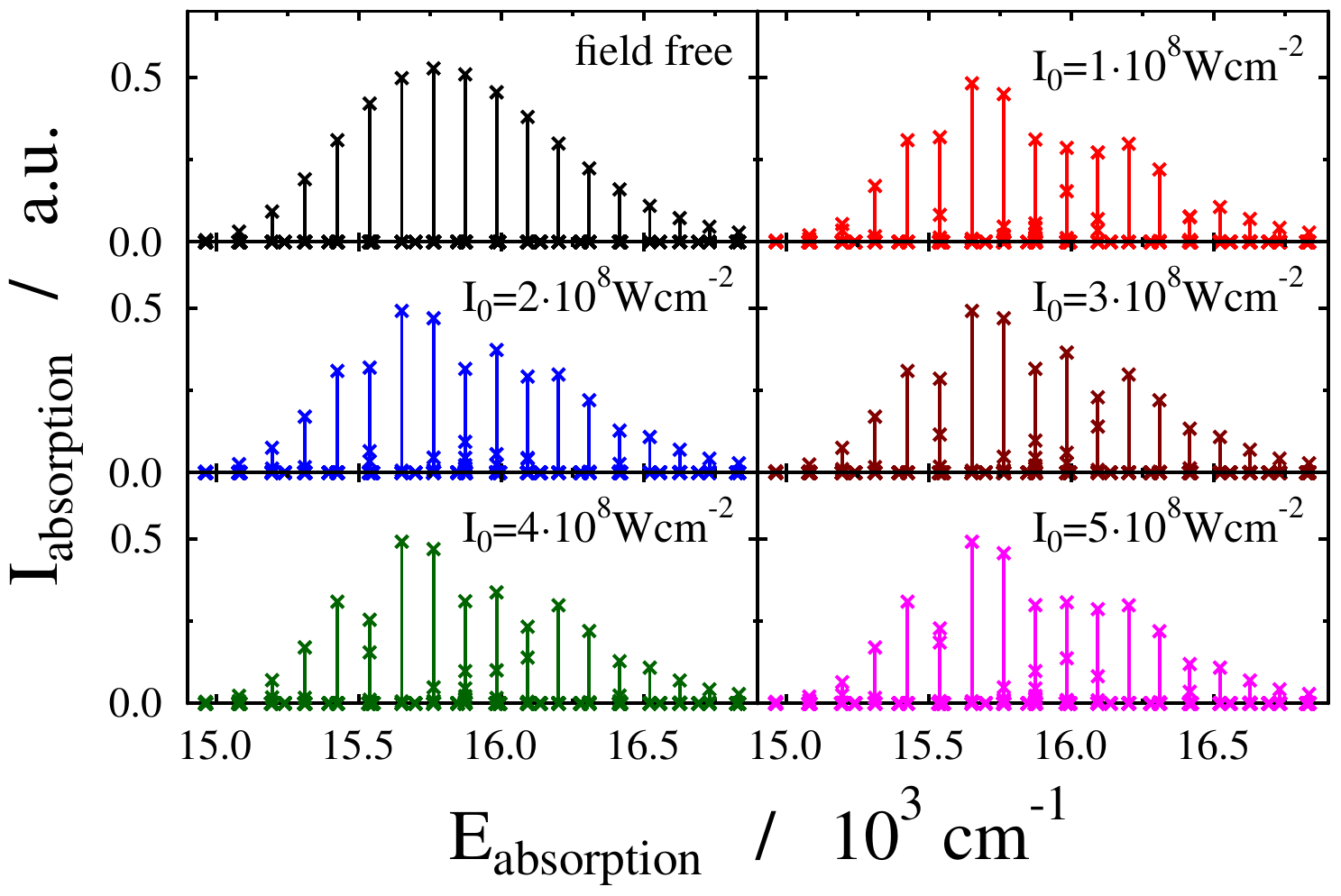}

\includegraphics[width=0.48\textwidth]{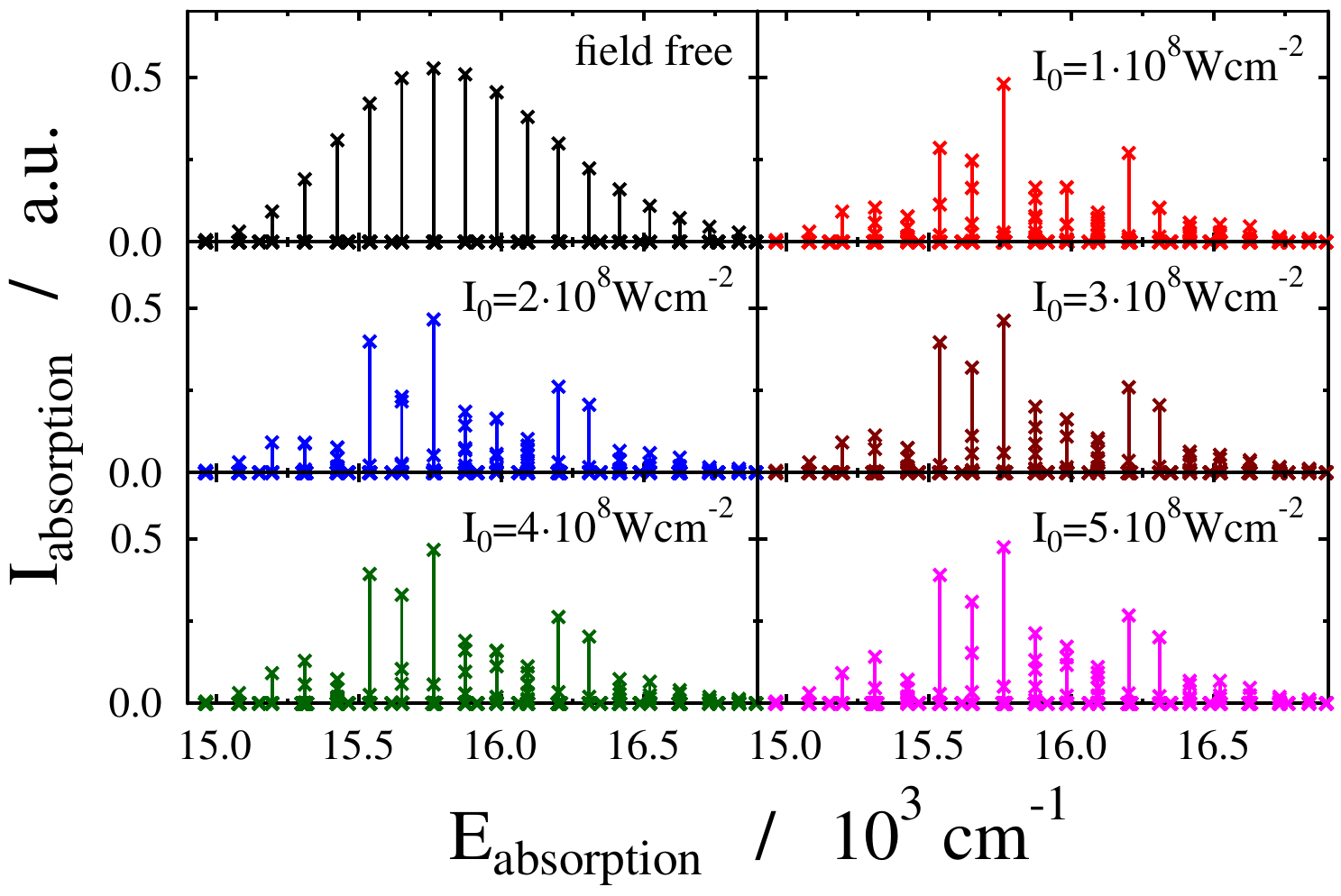}\hfill{}\includegraphics[width=0.48\textwidth]{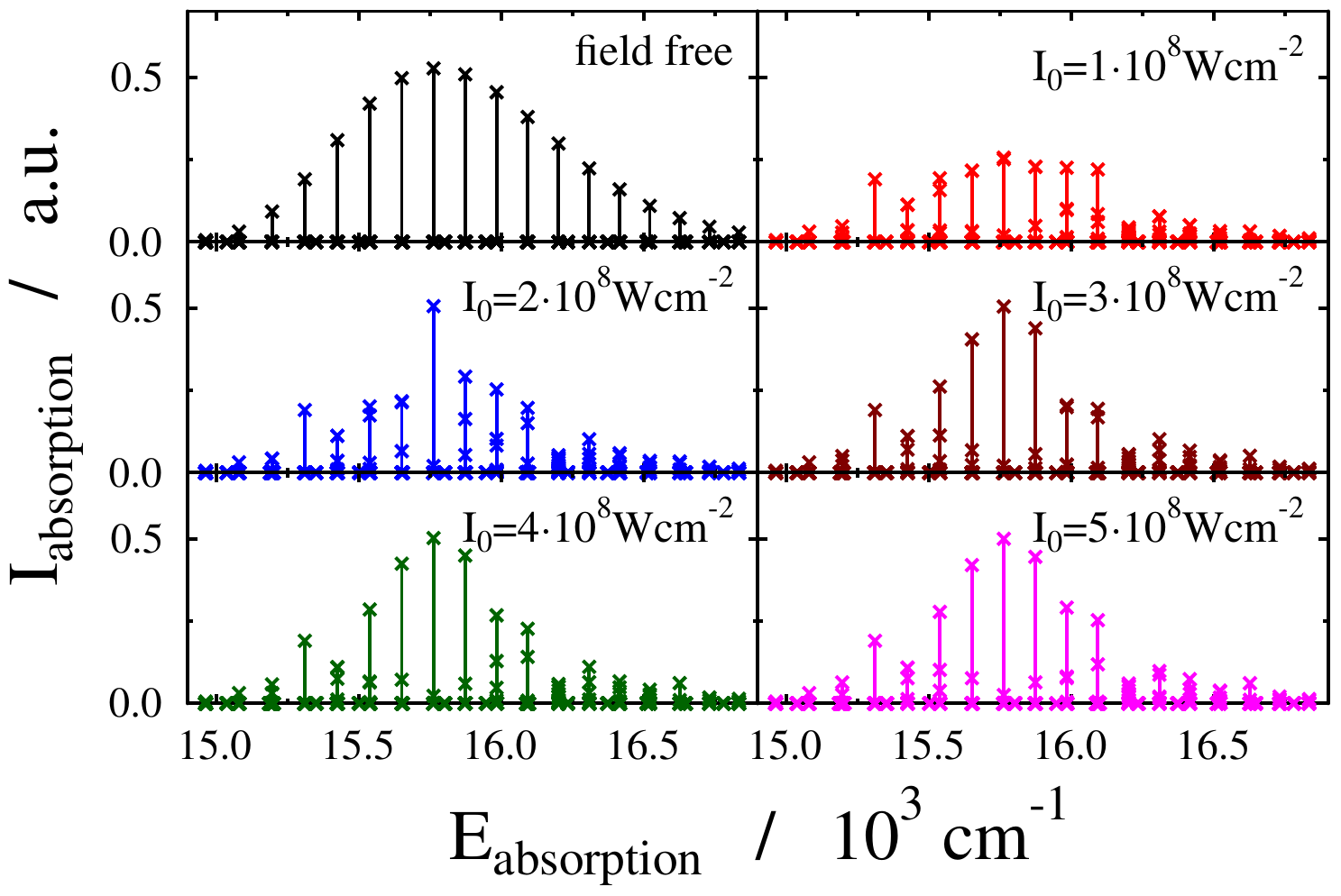}

\caption{\label{fig:6} Absorption spectra of the standing field-dressed Na$_{2}$
as a function of dressing-field intensity, computed using the first
expression of Eq \ref{eq:transition}. The spectra only show transitions
from a single field-dressed state, which was chosen by the ``second
adiabatic approximation''. In the upper panels (\textgreek{l}$_{L}$
= 663 nm) the curves shown in the left side are obtained by convolving
the right-hand side stick spectra with a Gaussian function having
\textgreek{sv} = 50 cm$^{-1}$. In the lower panels stick spectra
are for \textgreek{l}$_{L}$ = 660 nm (left) and \textgreek{l}$_{L}$
= 665 nm (right). }
\end{figure*}

\begin{figure}
\includegraphics[width=0.48\textwidth]{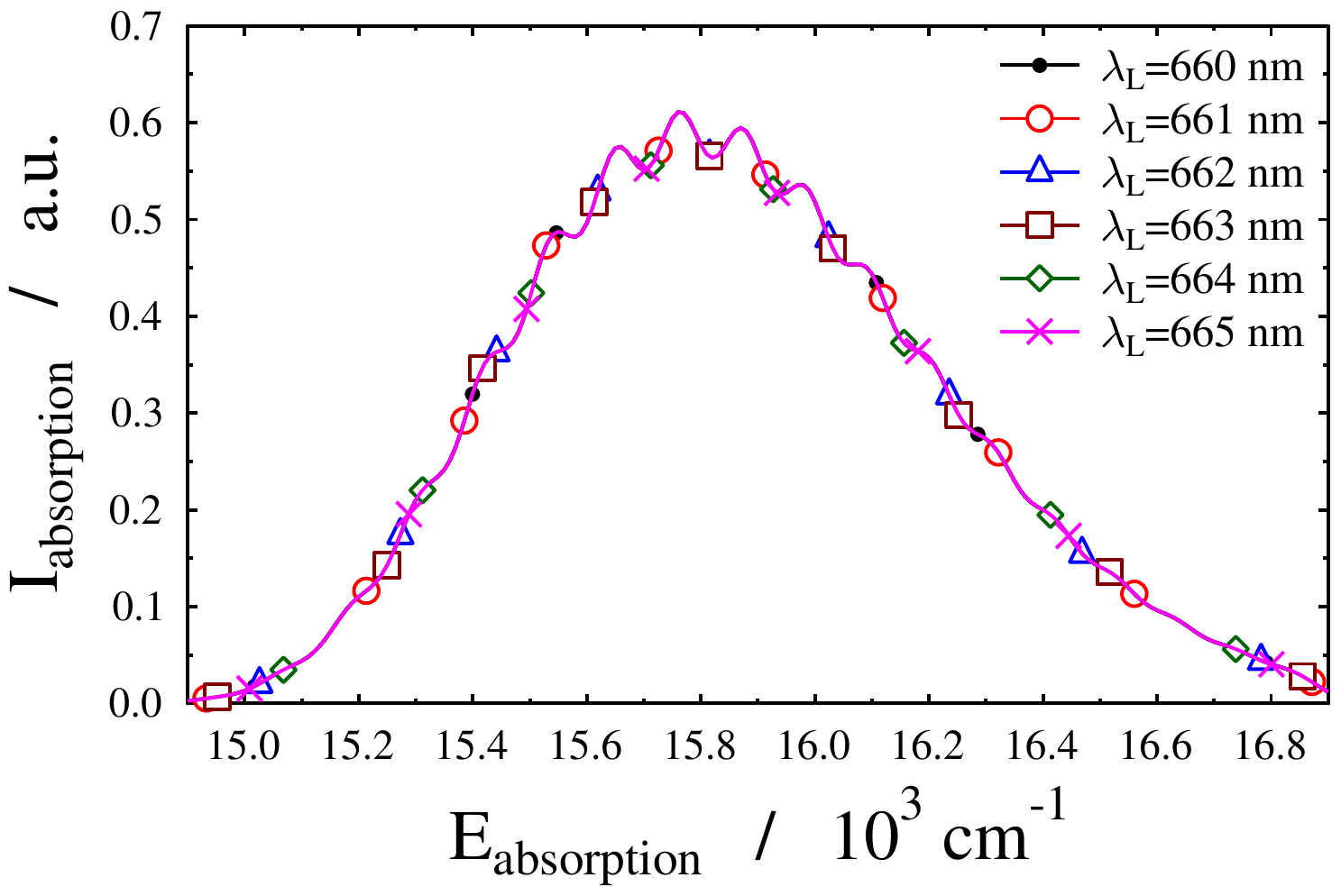}

\caption{\label{fig:7} Absorption spectra of the standing field-dressed Na$_{2}$
at six different dressing-field wavelengths and a dressing-field intensity
of I$_{0}$ =5 \texttimes{} 10$^{8}$ W/cm$^{2}$. The spectra only
show transitions from a single field-dressed state, which was chosen
by the ``second adiabatic approximation''.}
\end{figure}

\begin{figure}
\includegraphics[width=0.48\textwidth]{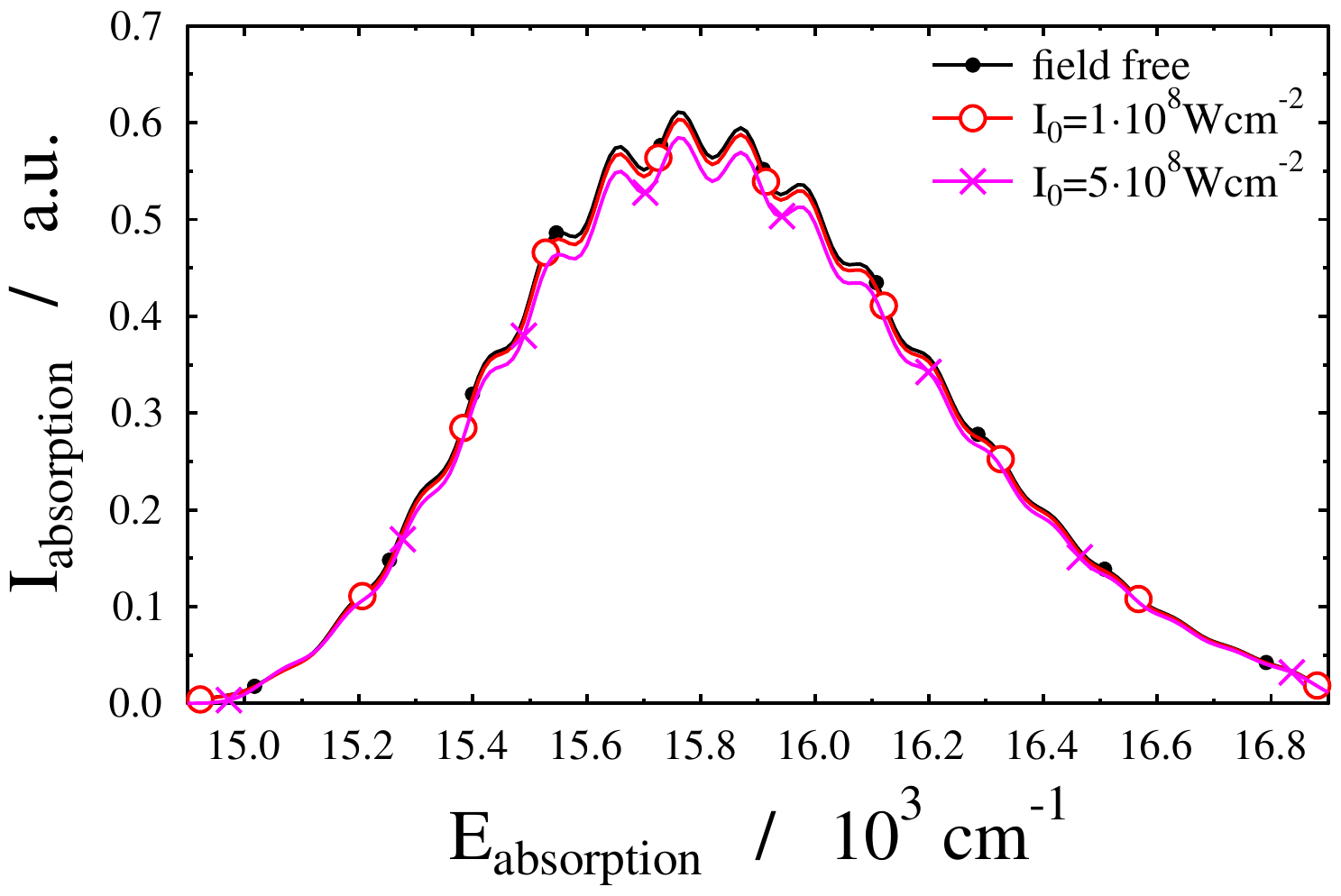}

\includegraphics[width=0.48\textwidth]{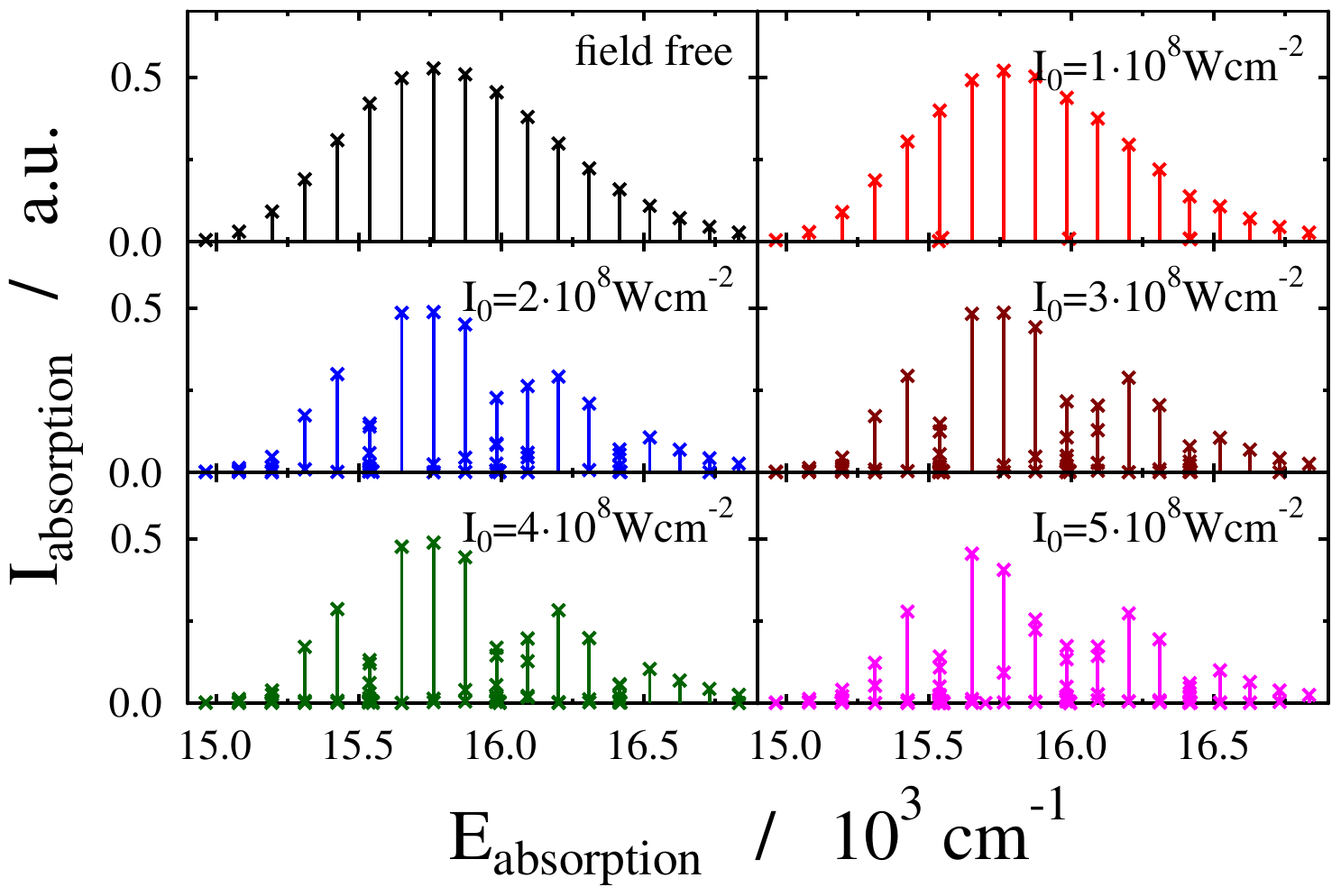}

\includegraphics[width=0.48\textwidth]{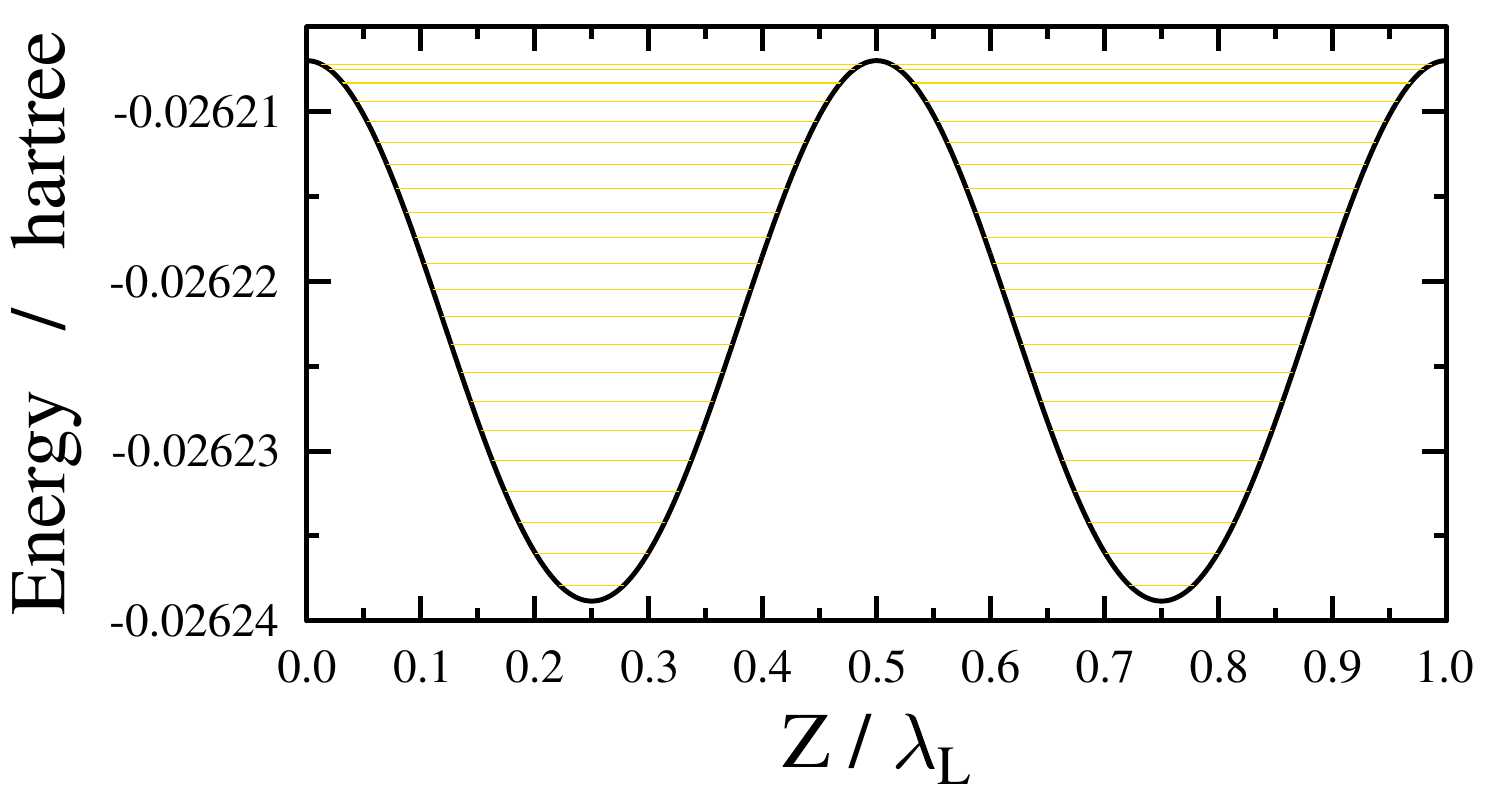}

\caption{\label{fig:8}Top panel: absorption spectra of the standing field-dressed
model based system (the mass is obtained by reducing the mass of the
Na$_{2}$ molecule to its 20000 portion) at different dressing-field
intensities ($\lambda_{L}=663$ nm). The spectra only show transitions
from a single field-dressed state, which was chosen by the ``second
adiabatic approximation''. The curves shown in the top panel are
obtained by convolving the stick spectra in the middle panel with
a Gaussian function having \textgreek{sv} = 50 cm$^{-1}$. Bottom
panel: translational energy levels of a model system wherein the actual
mass is obtained by reducing the mass of the Na$_{2}$ molecule to
its 20000 portion and I$_{0}$ =5 \texttimes{} 10$^{8}$ W/cm$^{2}$.}
\end{figure}

\section{Results and discussion}

We discuss the field-dressed absorption spectrum of the $\mathrm{Na_{2}}$
molecule in an optical lattice obtained from an initial state chosen
to be a single $|\Psi_{q}^{{\rm FD}}\rangle$ field-dressed state. 

\subsection{Adiabatic approximation}

Let us first choose the $|\Psi_{q}^{{\rm FD}}\rangle$ field-dressed
state according to the so-called ``adiabatic approximation'' \cite{adiabatic_theorem}.
Within this framework the field-dressed state chosen is the one populated
adiabatically from the field-free ground electronic and vibrational
molecular state by slowly switching on the dressing field. This approach
has been used previously in a similar context \cite{Tamas1} and provides
a reasonable approximation as long as the dressing field is turned
on much slower than the characteristic timescales of the system. Because
we can not suggest a specific $k$ value (see Eq. (\ref{eq:basis3}))
for the field-free initial state of an optical lattice experiment,
we identified the adiabatically populated field-dressed state as that
having the largest $\left|c_{q;0,k}^{(1)}\right|$ coefficient, see
Eq. (\ref{eq:field-dressed}), for a single but any $k$ value. In
Figs. \ref{fig:2} and \ref{fig:3} the field-dressed absorption spectra
are presented for several different dressing-light intensities and
wavelengths from $1\times10^{8}$ to $5\times10^{8}$ W/cm$^{2}$
and $660$ to $665$ nm. Inspecting Fig. \ref{fig:2} we see that
for the particular dressing-field wavelength of $663$ nm, the presence
of the dressing field leads to a decrease in the absorption of the
molecule with respect to the field-free case. In addition, tiny discrepancies
between the different field-dressed absorption envelopes can be realized.
Similarly, the absorption envelope obtained by a $663$ nm dressing
wavelength (see on Fig. \ref{fig:3}) also differs from the other
wavelength dressing fields.

These findings are not surprising, because such and even more pronounced
deviations have already been visualized on the spectra of freely rotating
$\mathrm{Na_{2}}$ molecules dressed by running wave laser lights
\cite{Tamas1}. For the sake of comparison we present in Fig. \ref{fig:4}
some field-dressed absorption spectra of the freely rotating $\mathrm{Na_{2}}$
for several different light intensities of a $663$ nm wavelength
dressing-field \footnote{For the calculation of the spectra the first term of the last line
in Eq. (3) of \cite{Tamas1} is applied.}. It can clearly be seen that the field-dressed absorption envelopes
are undergoing some remarkable changes via modifying the dressing-field
intensities. They can move either up or down relative to the position
of the field-free envelope.

The lower panels of Figs. \ref{fig:2} and \ref{fig:4}, showing the
individual spectral lines of the respective field-dressed spectra,
demonstrate significant differences between a non-rotating Na$_{2}$
dressed by an optical lattice and a freely rotating Na$_{2}$ dressed
by a running wave laser fields. In both cases Autler\textendash Townes-type
splittings \cite{AutlerTownes1} of the field-free peaks can be observed,
but the splittings are much more pronounced in the optical lattice
case. This is probably related to the fact that the density of translational
states in the optical lattice is much higher than the density of rotational
states for the freely-rotating Na$_{2}$, therefore, the mixing of
the different field-free translational states can occur more efficiently
than the mixing of the different rotational states during the dressing
process.

On the other hand, the spectrum envelope for the freely-rotating Na$_{2}$
is more sensitive to the dressing field intensity (see upper panels
of Figs. \ref{fig:2} and \ref{fig:4}). The changes in the overall
spectral peak heights resulting from couplings (induced by the dressing
field) between eigenstates of a zeroth-order (field-free) Hamiltonian
is referred to as intensity-borrowing \cite{Lenz1,Spect1}. It is
known that in classical vibronic molecular spectroscopy the intensity-borrowing
effect is a clear fingerprint of the presence of a strong nonadibatic
mixing between the different degrees of freedom in the close vicinity
of an intrinsic conical intersection. It has been shown that similar
situation holds for the dressed state spectra of a freely rotating
$\mathrm{Na_{2}}$ molecule when running laser waves are used for
the dressing resulting in the formation of a light-induced conical
intersection \cite{Tamas1}. In the optical lattice, the different
degree of intesity borrowing is most probably related to the fact
that while the transition amplitude induced by the probe pulse is
a function of the rotational $\theta$ coordinate, it is not a function
of the translational $Z$ coordinate, see Eq. (3) in Ref. \cite{Tamas1}
and Eq. (\ref{eq:transition}). Therefore, the mixing of the different
translational states in the light-dressed states can lead to slight
peak splittings, but do not lead to significant changes in the summed
overall transition amplitudes of Eq. (\ref{eq:transition}). Therefore,
the impact of the light-induced conical intersection is not so pronounced. 

As already stated before, during the above simulations an adiabatic
turn-on of the dressing field was assumed, \textit{i.e.}, the field-free
ground state was assumed to change into a single light-dressed state
in an adiabatic fashion. This light-dressed state was used as the
initial state when computing the field-dressed spectra. However, such
an adiabatic approximation is only valid if the turn-on time of the
dressing field is much longer than the characteristic timescales of
the system under study. For the freely rotating Na$_{2}$ this can
be a reasonable approximation, as the smallest rotational transition
energy of approximately 0.3 cm$^{-1}$ stands for a timescale of around
100 ps. On the other hand, the timescale corresponding to the very
dense, nearly continuous energy levels of the translational motion
in an optical lattice (see Fig. \ref{fig:5}) is very long, around
a few seconds. Such a slow turn-on time for the dressing field is
not realistic, thus the adiabatic approximation seems to be not so
adequate for the optical lattice case. 

\subsection{Second adiabatic approximation}

Another possible approach to select the initial light-dressed state
is to choose the one in which the sum of the field-free basis coefficients
$\sum_{k}\left|c_{q;\nu=0,k}^{(1)}\right|^{2}$ is the largest, i.e.
the field-dressed eigenvector with the largest contribution from the
field-free $\varphi_{\nu=0}^{(1)}$ is selected (see Eq. (\ref{eq:field-dressed})).
Here k numbers the basis functions related to the translational motion
(see Eq. (\ref{eq:basis3})). With such a selection process, we aim
to represent a physical scenario in which the turn-on time is adiabatic
with respect to the timescale of the vibrational and electronic degrees
of freedom, but instantaneous with respect to the timescale of the
translational degree of freedom. This selection criteria was applied
during the investigation of the molecular localization in an optical
lattice as well \cite{Pawlak} (we shortly refer to it as ``second
adiabatic approach''). Curves of the field-dressed absorption spectra
for several different light intensity and wavelength values obtained
by this framework are displayed in Figs. \ref{fig:6} and \ref{fig:7}.
The most striking result is the robustness of the spectra. We obtained
a significantly different picture than for the situation of the previously
studied freely-rotating Na$_{2}$ molecule. The absorption envelopes
are almost identical, not only at different intensities but also when
compared with the field-free one. In contrast, the height of the peaks
of the stick spectra are substantially different and each of which
differ prominently from the height of the field-free peaks. In other
words, new transitions appear due to the standing laser field formed
dressed states. Here again, the densely packed small pikes can be
understood as the fingerprint of the Autler\textendash Townes effect.
The presence of the huge number of small spikes results in a rich
structure in the line spectra, despite the fact that the envelopes
do not change their shape. The latter fact, however, allows us to
suggest that the intensity borrowing effect \textendash{} which can
be seen as the clear fingerprint of the light-induced conical intersection
\textendash{} is missing now. Investigating in detail the initial
field-dressed states obtained with the \textquotedbl{}adiabatic approximation\textquotedbl{}
or the \textquotedbl{}second adiabatic approximation\textquotedbl{}
reveals that while in the \textquotedbl{}adiabatic approximation\textquotedbl{}
the overall contribution of the $\varphi_{\nu=0}^{(1)}$ state is
around 91.63\% (for $I_{0}=10^{8}$ Wcm$^{-2}$ and $\lambda_{L}=663$
nm), it is almost 99.98\% for the \textquotedbl{}second adiabatic
approximaition\textquotedbl{}. This suggests, that for the intensity
borrowing effect to be seen in the optical lattice setup represented
in our simulations, the initially prepared field-dressed state should
be somewhat contaminated by the excited electronic state manifold.
This can be rationalized by the previously mentioned fact that the
mixing of only translational states does not lead to significant changes
in the transition amplitudes induced by the probe pulse.

\subsection{Model system}

By scaling artificially the total mass (denoted by $M$ in Eq. \ref{eq:translational-energy})
of the Na$_{2}$ molecule in an optical lattice without changing the
reduce mass ($\mu$) related to the vibrational motion, an increased
separation of the translational energy levels can be achieved. Fig.
\ref{fig:8} shows the translational energy levels of such a model
system wherein the actual mass is obtained by reducing the mass of
the Na$_{2}$ molecule to its 20000 portion. The resulting object
shows a close similarity with the previously studied rotating Na$_{2}$
molecule in the sense that the rotational energy levels of the molecule
are much further apart from each other than the very dense, almost
continuous energy levels belonging to the unscaled translation motion
in the lattice. Studying the field-dressed spectra of this model system
with the scaled total mass, it can be realized that similarly to the
field-dressed spectra of the freely-rotating Na$_{2}$ molecule (see
in Fig. \ref{fig:4}) the absorption envelopes, even if not so significantly,
but differ from each other. Due to some numerical problems that emerge
during the construction of the model, we could manage a reduction
of the mass of the Na$_{2}$ molecule only to its 20000 portion, but
even in this case it is clearly visible that the envelopes belonging
to different intensity values noticeably differ from each other. The
robustness of the spectrum disappears, the absorption envelopes are
moving, although these modifications are much less sensitive to the
change of the wavelength and intensity of the dressing field than
those of the freely-rotating Na$_{2}$. Even though the ``second
adiabatic approach'' was applied for the selection of the field-dressed
state - which keeps the absorption envelopes of the unscaled Na$_{2}$
in an optical lattice fixed \textendash{} the envelopes of the absorption
spectra are still changed as a result of the artificial widening of
the translational energy level gaps. This means that besides the always
existsing Autler\textendash Townes-type splittings the intensity borrowing
slowly appears as well, representing the signature of the light-induced
nonadibatic phenomena. This effect can be traced back once again to
the contamination of the initial field-dressed state with the excited
electronic state manifold, contribution of the $\varphi_{\nu=0}^{(1)}$
state is 98.69\% (for $I_{0}=10^{8}$ Wcm$^{-2}$ and $\lambda_{L}=663$
nm).

Obtained results demonstrate that irrespective of the evaluation algorithm
used for the selection of the field-dressed molecular states, the
very dense energy level structure of the translational motion taking
place in an optical lattice play an essential role in the robustness
of the field-dressed spectrum.

\section{Conclusions}

Assuming a weak probe pulse, we simulated the standing laser field-dressed
absorption spectra of non-rotating Na$_{2}$ molecules located in
an optical lattice. The direction of the applied probe pulse was set
to be perpendicular to the direction of the dressing pulse and the
polarization direction of the two fields were set the same. In this
case the dependence of the translational coordinate Z in the probe
pulse does not appear. Applying the proposed algorithm of \cite{Pawlak}
for the selection of a single dressed state which is the most similar
to the field-free ground electronic and vibrational state of the molecule,
a robust field-dressed absorption spectrum for the Na$_{2}$ molecule
was obtained. Although, the present model of the optical lattice provides
almost identical envelopes of the absorption spectra in the studied
wavelength and intensity ranges, it seems that this robustness can
be influenced by changing the density of the states and the method
of selecting the initial field-dressed state, which is of course,
not a trivial decision. 

We think that the present results are likely to change if the probe
pulse is applied in a different direction, because changes may appear
in the picture due to the emerging translational coordinate (Z) dependence
of the probe pulse. Furthermore, by envolving an additional degree
of freedom in the probe stage might allow us to visualize the signature
of the light-induced nonadibatic phenomena in the standing field-dressed
spectra, which is hardly present in this model. A more accurate method
for the standing laser field-dressed spectroscopy can be constructed
if rotations are included or if one considers spectra obtained from
not a single field-dressed state, but from the combination of several
of them. To continue, we plan to go in this direction. However, it
is a great challenge, as one has to solve the time-dependent dynamical
Schrödinger equation for the pump stage so as to determine the contributions
of the different field-dressed states composing the initial wave function
measured by the proceeding probe process. 
\begin{acknowledgments}
This research was supported by the EU-funded Hungarian grant EFOP-3.6.2-16-2017-00005.
The authors are grateful to NKFIH for support (Grant K128396 and PD124623). 
\end{acknowledgments}

\end{document}